\documentclass[showpacs,amsmath,amssymb,aps,showkeys,floatfix,prd,a4paper]{revtex4}

\usepackage[dvips]{graphicx}
\usepackage{dcolumn}
\usepackage{bm}
\usepackage{epsfig}
\usepackage{amsfonts}
\usepackage{amssymb,amscd}

\def\lsim{\raise0.3ex\hbox{$<$\kern-0.75em\raise-1.1ex\hbox{$\sim$}}}

\def\gsim{\raise0.3ex\hbox{$>$\kern-0.75em\raise-1.1ex\hbox{$\sim$}}}

\newcommand{\be}{\begin{equation}}

\newcommand{\ee}{\end{equation}}

\def\beq{\begin{equation}}

\def\eeq{\end{equation}}

\def\beqa{\begin{eqnarray}}

\def\eeqa{\end{eqnarray}}

\newcommand{\ba}{\begin{eqnarray}}

\newcommand{\rr}{\mbox{\boldmath $r$}}

\newcommand{\rb}{\mbox{\boldmath $b$}}

\def\gappeq{\mathrel{\rlap {\raise.5ex\hbox{$>$}}

{\lower.5ex\hbox{$\sim$}}}}

\def\lappeq{\mathrel{\rlap{\raise.5ex\hbox{$<$}}

{\lower.5ex\hbox{$\sim$}}}}

\def\Toprel#1\over#2{\mathrel{\mathop{#2}\limits^{#1}}}

\begin{document}

\begin{flushright}
LU TP 15-XX\\
December 2015
\vskip1cm
\end{flushright}

\title{Double vector meson production in $\gamma \gamma$ interactions at hadronic colliders}
\author{V.P. Gon\c{c}alves $^{1,2}$,  B.D.  Moreira$^{3}$  and  F.S. Navarra$^3$}
\affiliation{$^1$ Department of Astronomy and Theoretical Physics, Lund University, SE-223 62 Lund, Sweden \\  
$^{2}$ High and Medium Energy Group, Instituto de F\'{\i}sica e Matem\'atica,  Universidade Federal de Pelotas\\
Caixa Postal 354,  96010-900, Pelotas, RS, Brazil.\\
$^3$Instituto de F\'{\i}sica, Universidade de S\~{a}o Paulo,
C.P. 66318,  05315-970 S\~{a}o Paulo, SP, Brazil\\
}

\begin{abstract}
In this paper we revisit the double vector meson production in  $\gamma \gamma$ interactions at heavy ion collisions and present, by the first time, predictions for the $\rho\rho$ and $J/\Psi J/\Psi$ production in proton -- nucleus and proton -- proton collisions. 
In order to obtain realistic predictions for rapidity distributions and total cross sections for the double vector production in ultra peripheral hadronic collisions we take into account of the description of $\gamma \gamma \rightarrow VV$ cross section at low energies as well as its behaviour at large energies,  associated to the gluonic interaction between the color dipoles. 
Our results demonstrate that the double $\rho$ production is dominated by the low energy behaviour of the $\gamma \gamma \rightarrow VV$ cross section. In contrast, for the double $J/\Psi$ production, the contribution associated to the description of the QCD dynamics at high energies contributes significantly, mainly  in $pp$ collisions. Predictions for the RHIC, LHC, FCC and CEPC - SPPC energies are shown.
\end{abstract}



\maketitle

\vspace{1cm}

In recent  years a series of experimental results from RHIC \cite{star,phenix}, Tevatron \cite{cdf} and 
LHC \cite{alice, alice2,lhcb,lhcb2,lhcb_ups,cms1,cms2,cms3,Atlas}  demonstrated that the study of photon - induced interactions in hadronic colliders is 
feasible and that it can be used to, among other things, improve our knowledge  on  the nuclear gluon  distribution  
\cite{gluon,gluon2,gluon3,Guzey,vicwerluiz}, on details of QCD dynamics 
\cite{vicmag_mesons1,outros_vicmag_mesons,vicmag_update,motyka_watt,Lappi,griep,bruno1,bruno2}, on the mechanism  of quarkonium production 
\cite{Schafer,mairon1,mairon2,cisek,bruno1,bruno2}, on the Odderon \cite{vicodd1,vicodd2} and on the photon flux of the proton \cite{vicgus1,vicgus2}. 
These data have stimulated the development of the theoretical description of these processes as well as the proposal of new forward detectors to be installed 
in the LHC \cite{ctpps,marek}. 

The basic idea in the photon-induced processes is that an ultra relativistic charged hadron (proton or nucleus) creates  strong electromagnetic fields. 
A photon stemming from the electromagnetic field of one of the two colliding hadrons can interact with one photon coming from the other hadron (photon - photon process) 
or it can interact directly with the other hadron (photon - hadron process) \cite{upc,epa}. In these processes the total cross section  can be factorized in 
terms of the flux of equivalent  photons from the hadron projectile and the photon-photon or photon-target production cross section. In this paper we will focus on  
two -- photon interactions in hadronic collisions. Experimental results on  exclusive two-photon production of $W^+W^-$ and $\ell^+\ell^-$ pairs in 
$\gamma\gamma$ interactions reported by the CMS and ATLAS Collaborations \cite{cms1,cms2,cms3,Atlas} have demonstrated that it is possible to measure such events 
with the  experimental apparatus already available at the LHC, allowing for novel studies of QCD at very high energies and searches for Beyond Standard Model 
Physics (See, e.g., Ref. \cite{newphysics}). This  motivates us to revisit the analysis of  double vector meson production in ultra peripheral heavy ion 
collisions performed some time ago in Refs. \cite{vicmagvv1,vicmagvv2,vicmagvv3}, taking into account recent improvements in the description of the 
$\gamma \gamma \rightarrow VV$ ($V = \rho, J/\Psi$) cross section  at low \cite{antonirhorho,antonipsipsi} and at high \cite{brunodouble} energies. 
In  this work we will derive, for the first time,  realistic predictions for  double meson production in $\gamma \gamma$ interactions at $pp$ and $pA$ collisions 
at RHIC and  LHC energies as well as for the proposed energies  for the  Future Circular Collider (FCC) at CERN  \cite{fcc} and in the Circular Electron Positron 
Collider with a subsequent Super proton -- proton Collider  (CEPC - SPPC) in China \cite{cepc}. Our goal is to determine if this process, after the inclusion of the 
recent theoretical improvements, can be used to study the QCD dynamics at high energies, as originally proposed in Ref. \cite{vicmagvv1}. As we will show, this remains 
true for  double  $J/\Psi$ production, mainly in $pp$ collisions.

Let us start our analysis presenting a brief review of the main formulas to describe double vector meson production in $\gamma \gamma$ interactions at 
hadronic colliders. In the Equivalent Photon Approximation (EPA) \cite{upc,epa} the total cross section for this process can be written as
\begin{eqnarray}
\sigma \left( h_1 h_2 \rightarrow h_1 \otimes V_1V_2 \otimes h_2 ;s \right)   
&=& \int \hat{\sigma}\left(\gamma \gamma \rightarrow V_1V_2 ; 
W \right )  N\left(\omega_{1},{\mathbf b_{1}}  \right )
 N\left(\omega_{2},{\mathbf b_{2}}  \right ) S^2_{abs}({\mathbf b})  
 \mbox{d}^{2} {\mathbf b_{1}}
\mbox{d}^{2} {\mathbf b_{2}} 
\mbox{d} \omega_{1}
\mbox{d} \omega_{2} \,\,\, ,
\label{cross-sec-1}
\end{eqnarray}
where $\sqrt{s}$ is center - of - mass energy of the $h_1 h_2$ collision ($h_i = p, A$), $\otimes$ characterizes a rapidity gap in the final state and 
$W = \sqrt{4 \omega_1 \omega_2}$ is the invariant mass of the $\gamma \gamma$ system. Moreover, $N(\omega, {\mathbf b})$ is the equivalent photon spectrum  
of photons with energy $\omega$ at a distance ${\mathbf b}$  from the hadron trajectory, defined in the plane transverse to the trajectory. The spectrum can be 
expressed in terms of the charge form factor $F$ as follows
\begin{eqnarray}
 N(\omega,b) = \frac{Z^{2}\alpha_{em}}{\pi^2}\frac{1}{b^{2}\omega}
\cdot \left[
\int u^{2} J_{1}(u) 
F\left(
 \sqrt{\frac{\left( \frac{b\omega}{\gamma_L}\right)^{2} + u^{2}}{b^{2}}}
 \right )
\frac{1}{\left(\frac{b\omega}{\gamma_L}\right)^{2} + u^{2}} \mbox{d}u
\right]^{2} \,\,,
\label{fluxo}
\end{eqnarray}
where $\gamma_L$ is the Lorentz factor. The factor $S^2_{abs}({\mathbf b})$ is the absorption factor, given in what follows by
\begin{eqnarray}
S^2_{abs}({\mathbf b}) = \Theta\left(
\left|{\mathbf b}\right| - R_{h_1} - R_{h_2}
 \right )  = 
\Theta\left(
\left|{\mathbf b_{1}} - {\mathbf b_{2}}  \right| - R_{h_1} - R_{h_2}
 \right )  \,\,,
\label{abs}
\end{eqnarray}
where $R_{h_i}$ is the radius of the hadron $h_i$ ($i = 1,2$). 
The presence of this factor in Eq. (\ref{cross-sec-1})  excludes the overlap between the colliding hadrons and allows to take into account only ultra 
peripheral collisions. Remembering that the photon energies $\omega_1$ and $\omega_2$  are related to   
$W$ and the rapidity ( $Y = \frac{1}{2}(y_{V_1} + y_{V_2})$ ) of the outgoing double meson system by 
\begin{eqnarray}
\omega_1 = \frac{W}{2} e^Y \,\,\,\,\mbox{and}\,\,\,\,\omega_2 = \frac{W}{2} e^{-Y} \,\,\,
\label{ome}
\end{eqnarray}
the total cross section can be expressed by (For details see e.g. Ref. \cite{mariola})
\begin{eqnarray}
\sigma \left( h_1 h_2 \rightarrow h_1 \otimes V_1V_2 \otimes h_2 ;s \right)   
&=& \int \hat{\sigma}\left(\gamma \gamma \rightarrow V_1V_2 ; 
W \right )  N\left(\omega_{1},{\mathbf b_{1}}  \right )
 N\left(\omega_{2},{\mathbf b_{2}}  \right ) S^2_{abs}({\mathbf b})  
\frac{W}{2} \mbox{d}^{2} {\mathbf b_{1}}
\mbox{d}^{2} {\mathbf b_{2}} 
\mbox{d}W 
\mbox{d}Y \,\,\, .
\label{cross-sec-2}
\end{eqnarray}
It is important to emphasize that in EPA we disregard the photon virtualities, which is a good approximation, mainly for ions, since the typical virtualities are 
$< 1/R_h$. Moreover, the highest energy of the photons is of the order of the inverse Lorentz contracted radius of the hadron 
$\approx \gamma_L/R_h$, with the spectra decreasing exponentially at larger energies. Consequently, for the same Lorentz factor, we have 
$W_{max}^{pp} > W_{max}^{pA} > W_{max}^{AA}$. Finally, due to the $Z^2$ dependence of the photon spectra,  for a fixed $W$ the following hierarchy 
is valid for processes induced by $\gamma \gamma$ interactions: $\sigma_{AA} \sim Z^2 \cdot \sigma_{pA} \sim Z^4 \cdot \sigma_{pp}$.

In order to estimate this cross section we must  describe the $\gamma \gamma \rightarrow V_1V_2$ interaction in a large   energy range. 
In what follows we will assume that
\begin{eqnarray}
\hat{\sigma}\left(\gamma \gamma \rightarrow V_1V_2 ; 
W \right ) = \hat{\sigma}^{LE}\left(W \right ) + \hat{\sigma}^{HE}\left(W \right )
\end{eqnarray}
where the $LE$ term is associated to the description of the cross section at low energies $W \lesssim 10$ GeV, while the $HE$ term describes the region of larger 
values of  $W$. Double vector meson production at low energies has been discussed and improved in Refs. \cite{antonirhorho,antonipsipsi}. As in 
Ref. \cite{antonirhorho},  we will evaluate  double $\rho$ production in $\gamma \gamma$ interactions directly from the experimental measurements using a fit to the world 
data, which describes the experimental data in the region of few GeV. In particular, we will take into account the huge enhancement close to the threshold observed 
in the data, which is not yet well understood. As demonstrated in Ref.  \cite{antonirhorho} this contribution determines the behavior of  double $\rho$ 
production in $AA$ collisions. In the case of  double $J/\Psi$ production, as in Ref. \cite{antonipsipsi}, we will consider the contribution associated to the 
box  diagrams, calculated in the heavy quark non - relativistic approximation. In Ref. \cite{antonipsipsi} the authors  have also estimated the contribution associated 
to the two - gluon exchange, which implies a $\gamma \gamma \rightarrow J/\Psi J/\Psi$ cross section independent of the energy. In our analysis, we will not 
include this contribution in  the low enegy term, since it is the leading order term in the dipole - dipole interaction present in our formalism to treat the 
high energy term discussed in what follows. Finally, it is important to emphasize that one of the main conclusions from Ref. \cite{antonipsipsi} is that in PbPb
collisions the box mechanism significantly dominates over the two - gluon exchange one.

\begin{figure}
\centerline{\psfig{figure=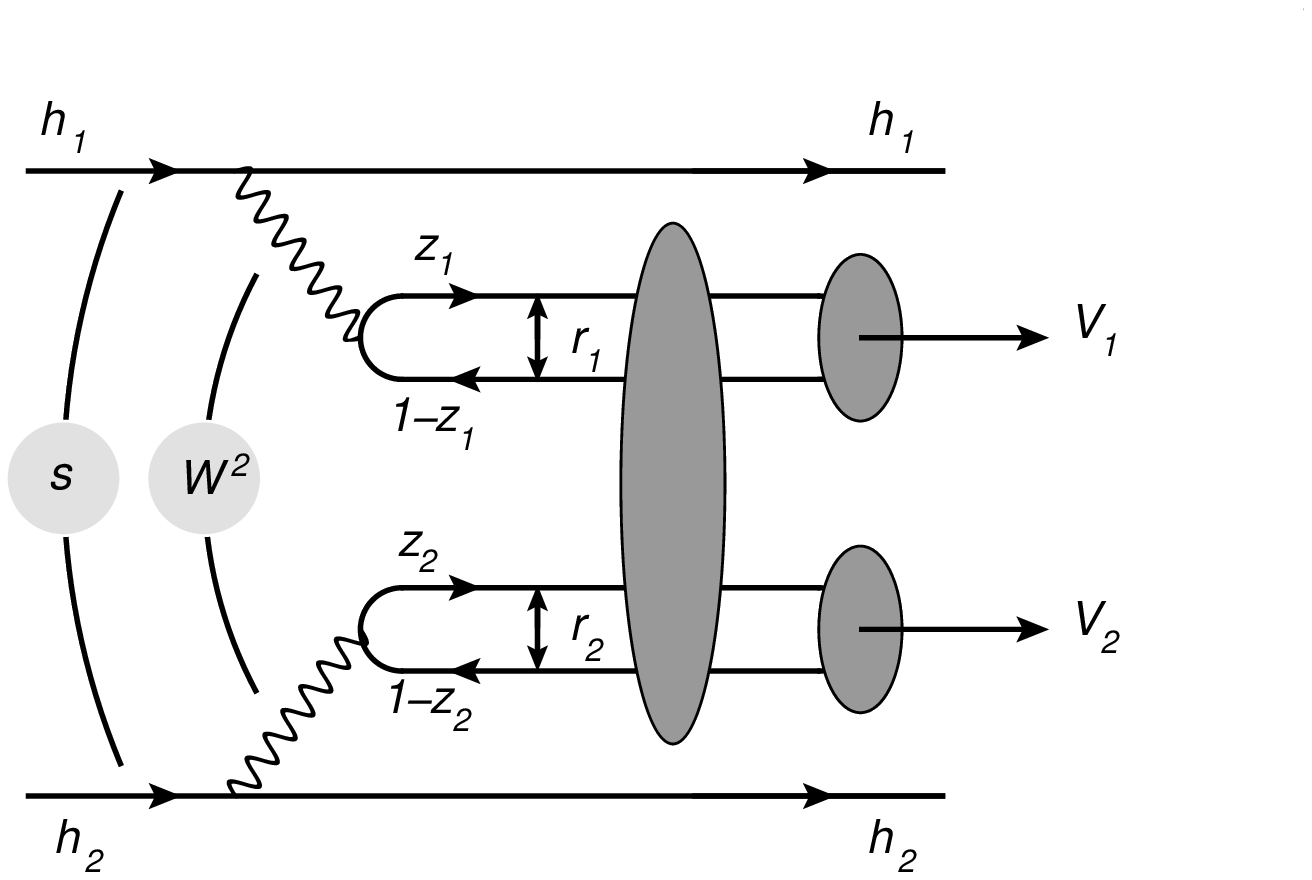,width=10cm}}
\caption{Double vector meson production in  $\gamma \gamma$ interactions at hadronic colliders  in the 
color dipole picture.}
\label{fig1}
\end{figure}

The description of   double vector meson production in $\gamma \gamma$ interactions at high energies  has attracted the attention of several theoretical groups 
in the last years, with the cross section being estimated in different theoretical frameworks 
\cite{serbo2,motyka,Qiao,dmvic1,dmvic2,Pire,vicmag07,Ivanov,antonirhorho,antonipsipsi,brunodouble}, as, for instance, the solution of the BFKL equation and 
impact factors at leading and next-to-leading orders.  
In particular,  in Ref. \cite{brunodouble} we have estimated the total $\gamma (Q_1^2) + \gamma (Q_2^2) \rightarrow V_1 + V_2$ cross-sections  for 
$V_i = \rho$, $\phi$, $J/\psi$ and $\Upsilon$ in the color dipole formalism considering the improved treatment of the dipole - dipole cross section proposed 
in Ref. \cite{nosfofo}. We have also taken advantage of the progress in the  knowledge of vector meson wave functions. 
Moreover, in Ref. \cite{brunodouble} we have  taken into account of the non-linear effects in the QCD dynamics, which are expected to be present at large energies. 
An important aspect of the analysis presented in Ref. \cite{brunodouble} is that the main ingredients are constrained by  LEP and HERA data. In particular, 
assuming the values for the slope parameter $B_{V_1V_2}$ proposed in Ref. \cite{vicmag07}, it is possible to obtain parameter-free predictions for the 
$\rho \rho $ and  $J/\Psi J/\Psi$ production cross sections at high energies.
In the case of ultra peripheral hadronic collisions,   double meson production is induced by the interaction of real photons and can be represented by the diagram in 
Fig. \ref{fig1}  in the color dipole formalism. In this approach the  $\gamma \gamma \rightarrow V_1 V_2$ interaction at hadronic colliders   can be seen 
as a succession in time of four factorizable subprocesses (See Fig. \ref{fig1}): i) the photons are emitted by the incident hadrons, ii) the photons fluctuate 
into  
quark-antiquark pairs (the dipoles), iii) these color dipoles interact and, iv) the pairs convert into the vector meson final states. In particular, the 
$\gamma \gamma \rightarrow V_1 V_2$ cross section can be expressed as follows
\begin{eqnarray}
\sigma\, (\gamma \gamma \rightarrow V_1 \, V_2) = \int dt  \,  \frac{d\sigma  
(\gamma \gamma \rightarrow V_1 \, V_2)}{dt}\, =  \frac{1}{B_{V_1 \,V_2}} \left. \frac{d\sigma  
(\gamma \gamma \rightarrow V_1 \, V_2)}{dt}\,\right|_{t_{min}=0} =  \frac{[{\cal I}m \, {\cal A}(W^2,\,t=0)]^2}{16\pi\,B_{V_1 \,V_2}} \;,
\label{totalcs}
\end{eqnarray}
where we have  approximated the $t$-dependence of the differential cross section by an exponential  with  $B_{V_1 \, V_2}$ being 
the slope parameter. The imaginary part of the amplitude at zero momentum transfer ${\cal A}(W^2,\,t=0)$ reads as
\begin{eqnarray}
{\cal I}m \, {\cal A}\, (\gamma \gamma \rightarrow V_1 \, V_2) & = &  
\int dz_1\, d^2\rr_1 \,[\Psi^\gamma(z_1,\,\rr_1)\,\, \Psi^{V_1*}(z_1,\,\rr_1)]_T \nonumber \\
&\times & \int dz_2\, d^2\rr_2 \,[\Psi^\gamma(z_2,\,\rr_2)\,\, \Psi^{V_2 *}(z_2,\,\rr_2)]_T
\,
\sigma_{d d}(\rr_1, \rr_2,Y)
 \, ,
\label{sigmatot}
\end{eqnarray}
where $\Psi^{\gamma}$ and $\Psi^{V_i}$  are the light-cone wave functions  of the photon and vector meson, respectively, and $T$ the transverse polarization.  
The variable $\rr_1$ defines the relative transverse
separation of the pair (dipole) and $z_1$ $(1-z_1)$ is the longitudinal momentum fraction of the quark (antiquark). Similar definitions hold  
for $\rr_2$ and  $z_2$. {  The variable $Y$ is the rapidity and will be defined  later}. 
The basic blocks are the photon wave function, $\Psi^{\gamma}$, the  meson wave function, $\Psi^{V}$,  and the dipole-dipole  cross
section, $\sigma_{d\,d}$. In contrast to the photon wave function, which is well known in the literature (See e.g. \cite{KMW}), the description of the vector meson 
wave functions is still a subject of debate. The simplest approach  is to assume that the vector meson is predominantly a quark-antiquark state and that the spin and 
polarization structure is the same as in the  photon \cite{dgkp,nnpz,sandapen,KT}. As in Ref. \cite{brunodouble} we will assume that 
the overlap between the photon and the vector meson wave function, for the transversely polarized  
case, is given by (for details see Ref. \cite{KMW})
\begin{eqnarray}
(\Psi_{V}^* \Psi)_T = \hat{e}_f e \frac{N_c}{\pi z (1-z)}\left\{m_f^2K_0(\epsilon r)\phi_T(r,z) -[z^2+(1-z)^2]\epsilon K_1(\epsilon r) \partial_r \phi_T(r,z)\right\} \,\,,
\end{eqnarray}
where $ \hat{e}_f $ is the effective charge of the vector meson, $m_f$ is the quark mass, $N_c = 3$, $\epsilon^2 = z(1-z)Q^2 + m_f^2$ and $\phi_T(r,z)$ defines  the 
scalar part of the  vector meson wave function. In what follows we will consider the Gauss-LC model  for $\phi_T(r,z)$, which  is then  given by
\begin{eqnarray}
\phi_T(r,z) = N_T [z(1-z)]^2 \exp\left(-\frac{r^2}{2R_T^2}\right) \,\,.
\end{eqnarray}
The parameters $N_T$ and $R_T$ are  determined by the normalization condition of the wave function and by the decay width (See Ref. \cite{brunodouble} for details).
The other main input to calculate the $\gamma \gamma \rightarrow V_1 V_2$ cross section is the 
dipole-dipole  cross
section, $\sigma_{d\,d}$.  
At  lowest order, the dipole - dipole interaction can be described by the two - gluon exchange between the dipoles, with the 
resulting cross section being energy independent (See, e.g. Ref. \cite{Navelet}). The inclusion of the leading corrections 
associated to terms $\propto \log(1/x)$ (as described by the BFKL equation) leads to a power-law energy behavior of the cross 
section, which violates the unitarity at high energies. Unitarity corrections were introduced in Ref. \cite{salam}, 
considering the color dipole picture and  independent multiple scatterings between the dipoles. These corrections were also addressed  in Ref. \cite{iancu_mueller} 
in the context of  the Color Glass Condensate (CGC) formalism \cite{cgc}.

In the eikonal approximation the dipole - dipole cross section can be expressed as follows: 
\begin{eqnarray}
\sigma^{dd} (\rr_1,\rr_2,Y)  = 2 \int d^2\rb \,{\cal{N}}(\rr_1,\rr_2,\rb,Y)
\end{eqnarray}
where  ${\cal{N}}(\rr_1,\rr_2,\rb,Y)$  is the scattering amplitude of the two dipoles with transverse sizes $\rr_1$ and $\rr_2$, 
relative impact parameter $\rb$ and rapidity separation $Y$. 
The interaction of two dipoles of similar sizes  is still an open question (See, e.g. Ref. \cite{kov_dd}). In a first approximation, it is useful to 
express ${\cal{N}}$ in terms of the solution of the Balitsky -- Kovchegov (BK) equation    (obtained disregarding the $\rb$ dependence), which has been derived 
considering an  asymmetric frame where the projectile has a simple structure and the evolution occurs in the target wave function \cite{BK}. A shortcoming of this 
approach is that, although the unitarity of the $S$-matrix (${\cal{N}} \le 1$) is respected by the solution of the BK equation, the 
associated dipole - dipole cross section can still rise indefinitely with the energy, even after the black disk limit (${\cal{N}} = 1$) has been reached at 
central impact parameters, due to the non-locality of the evolution. In Ref.  \cite{nosfofo} we have proposed a more elaborated model for the impact parameter 
dependence  in order to obtain more realistic predictions for the dipole - dipole cross section. Basically, we assumed that only the range $b < R$, where 
$R =$ Max$(r_1,r_2)$, contributes to the dipole - dipole cross section, i.e. we assumed that ${\cal{N}}$ is negligibly small when the dipoles have no overlap with 
each other ($b>R$). Therefore the dipole-dipole cross section can be expressed as follows \cite{nosfofo}: 
\begin{eqnarray}
\sigma^{dd} (\rr_1,\rr_2,Y)  = 2 \, {{N}}(\rr,Y) \int_0^R d^2\rb = 2 \pi R^2 {{N}}(\rr,Y)\,\,, 
\label{geral}
\end{eqnarray}
where ${{N}}(\rr,Y)$ is the forward scattering amplitude, which can be obtained as a  solution of the BK equation disregarding the impact parameter dependence 
or from phenomenological models that describe the HERA data.  The explicit form of $\sigma^{dd}$ reads
\begin{eqnarray}
\sigma^{dd} (\rr_1,\rr_2,Y) =  2 \pi r_1^2 N(r_2,Y_2) \, \Theta(r_1 - r_2) + 2 \pi r_2^2 N(r_1,Y_1) \, \Theta(r_2 - r_1) \,\,,
\label{ourmodel}
\end{eqnarray}
where  $Y_i = \ln (1/x_i)$ and 
\begin{eqnarray}
 x_i = \frac{Q_i^2 + 4 m_f^2}{W^2 + Q_i^2}.
\label{xdef}
\end{eqnarray}

As in Refs. \cite{nosfofo,brunodouble} we will consider in our calculations the IIM-S model \cite{iim,soyez} for the forward scattering amplitude, which is based 
on the solutions of the BK equation at small and large dipoles, and is given by
\begin{eqnarray}
{{N}}(\rr,Y) =  \left\{ \begin{array}{ll} 
{\mathcal N}_0\, \left(\frac{r\, Q_s}{2}\right)^{2\left(\gamma_s + 
\frac{\ln (2/r Q_s)}{\kappa \,\lambda \,Y}\right)}\,, & \mbox{for $r 
Q_s(x) \le 2$}\,,\\
 1 - \exp^{-a\,\ln^2\,(b\,r\, Q_s)}\,,  & \mbox{for $r Q_s(x)  > 2$}\,, 
\end{array} \right.
\label{CGCfit}
\end{eqnarray}
where $a$ and $b$ are determined by continuity conditions at $\rr Q_s(x)=2$,  $\gamma_s= 0.6194$, $\kappa= 9.9$, $\lambda=0.2545$, 
$Q_0^2 = 1.0$ GeV$^2$,
$x_{0}=0.2131\times 10^{-4}$ and ${\mathcal N}_0=0.7$.
As demonstrated in Ref. \cite{nosfofo}, using this model we can describe the LEP data for the total $\gamma \gamma$ cross sections and photon structure functions.

\begin{figure}
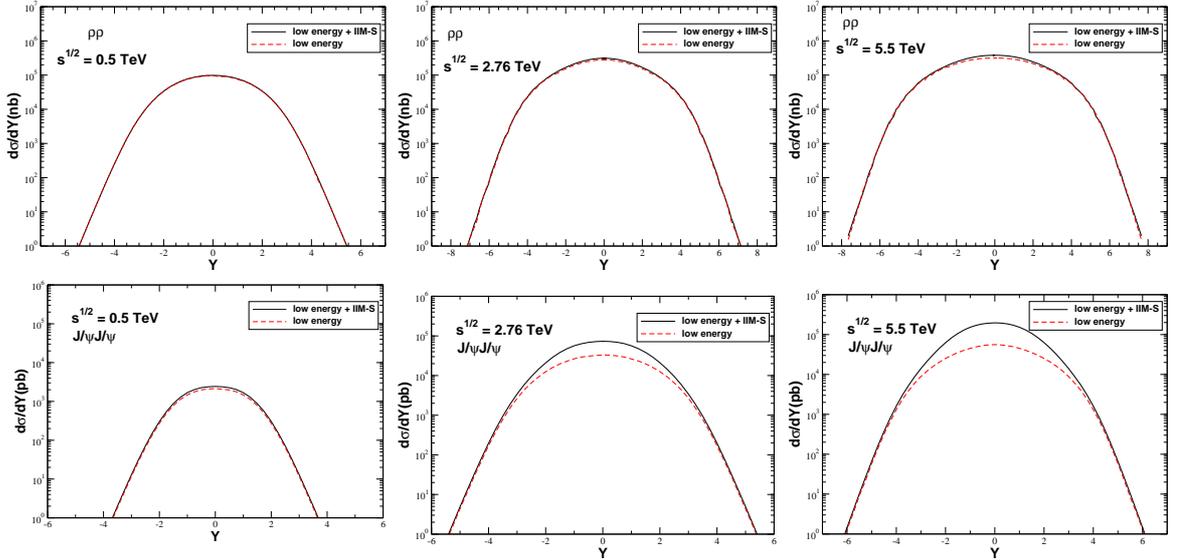

\begin{tabular}{ccc}
{\psfig{figure=AA-rho-rho-500.eps,width=5cm}} & 
{\psfig{figure=AA-rho-rho-2760.eps,width=5cm}} & 
{\psfig{figure=AA-rho-rho-5500.eps,width=5cm}} \\
{\psfig{figure=AA-psi-psi-500.eps,width=5cm}} & 
{\psfig{figure=AA-psi-psi-2760.eps,width=5cm}} & 
{\psfig{figure=AA-psi-psi-5500.eps,width=5cm}}
\end{tabular}                                                                                                                       
\caption{Rapidity distribution in double vector meson production in $\gamma \gamma$ interactions at $PbPb$ collisions considering different
 values of $\sqrt{s}$.}
\label{fig2}
\end{figure}

\begin{figure}
\begin{tabular}{cc}
{\psfig{figure=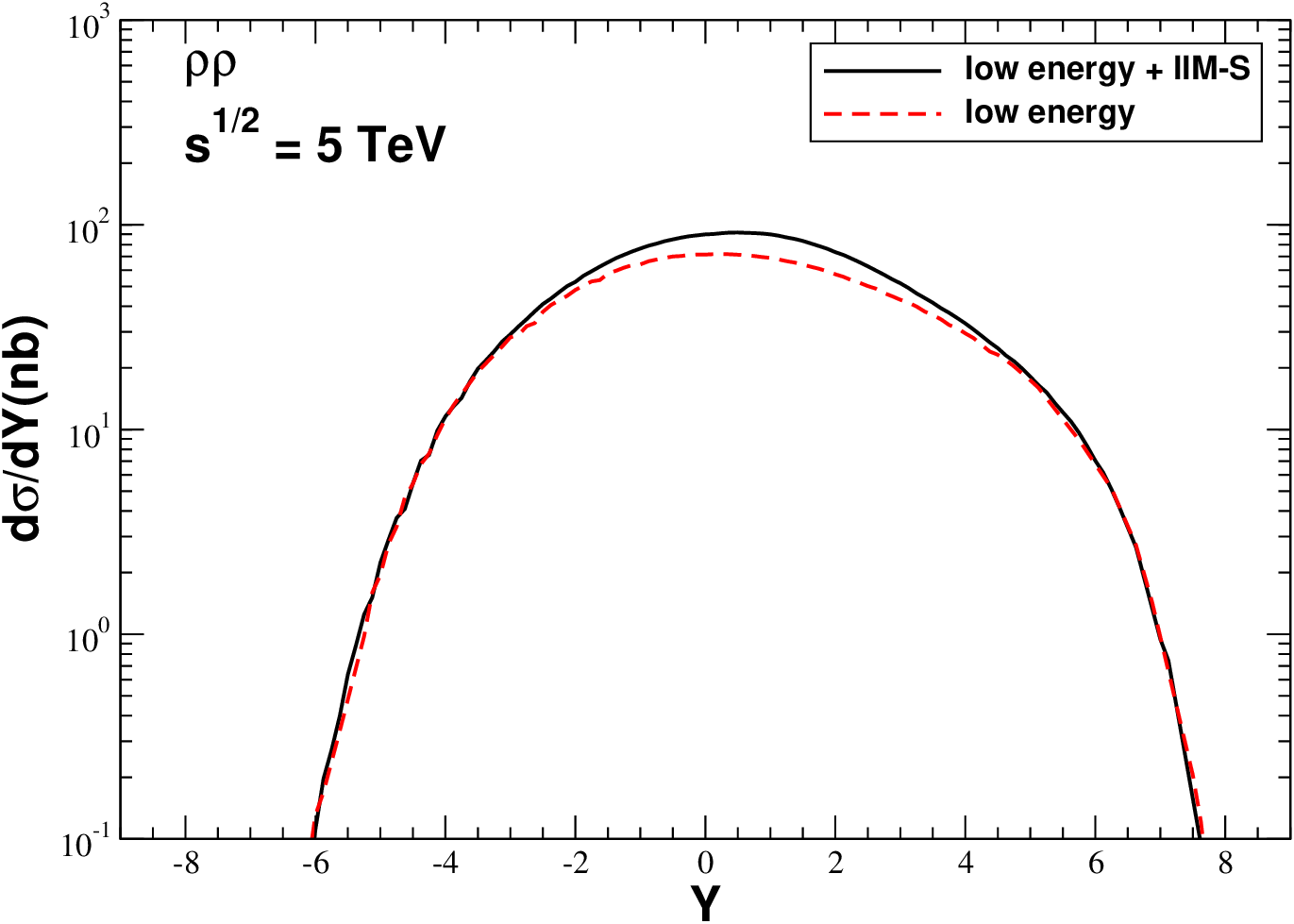,width=7cm}} & 
{\psfig{figure=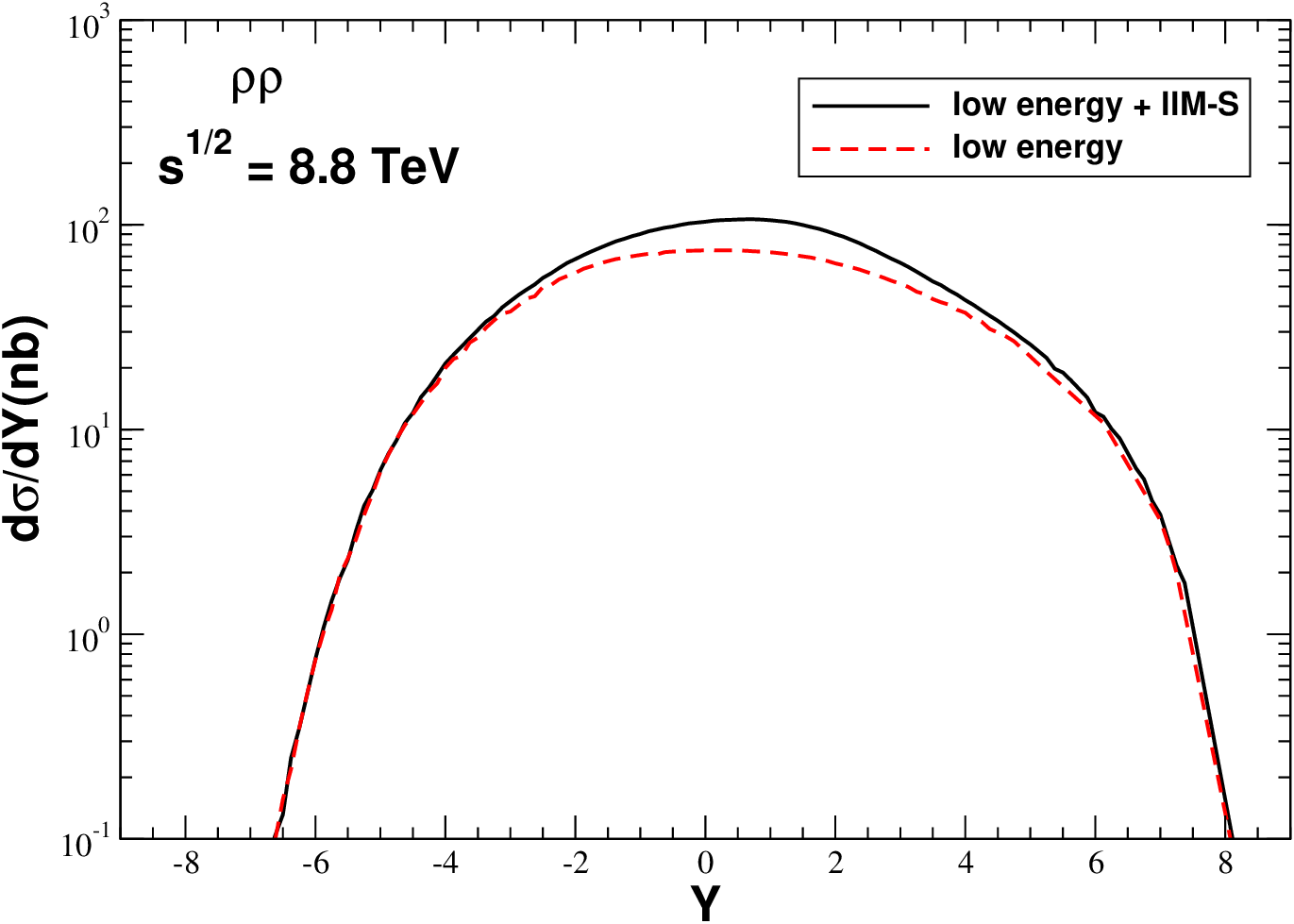,width=7cm}} \\
{\psfig{figure=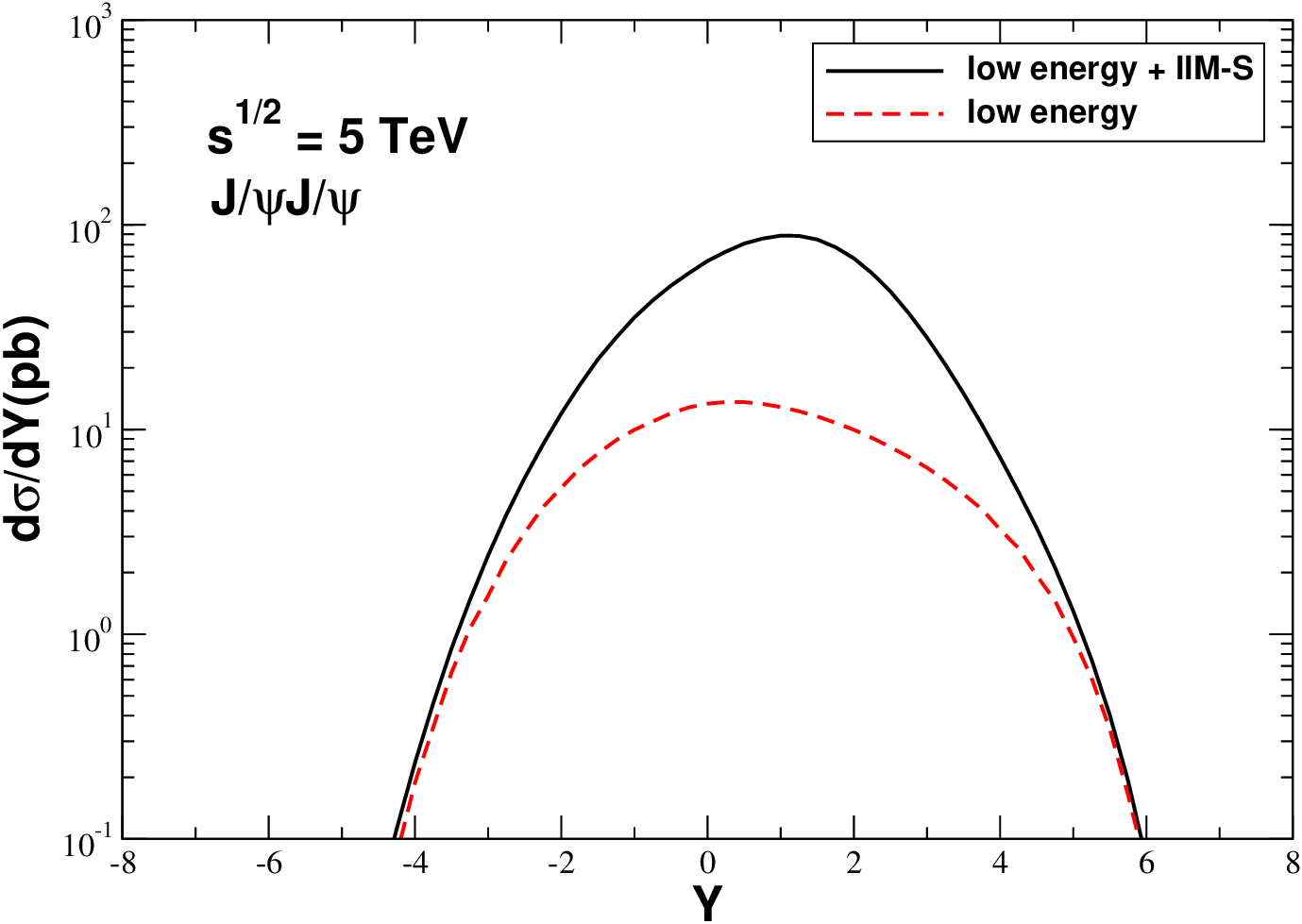,width=7cm}} & 
{\psfig{figure=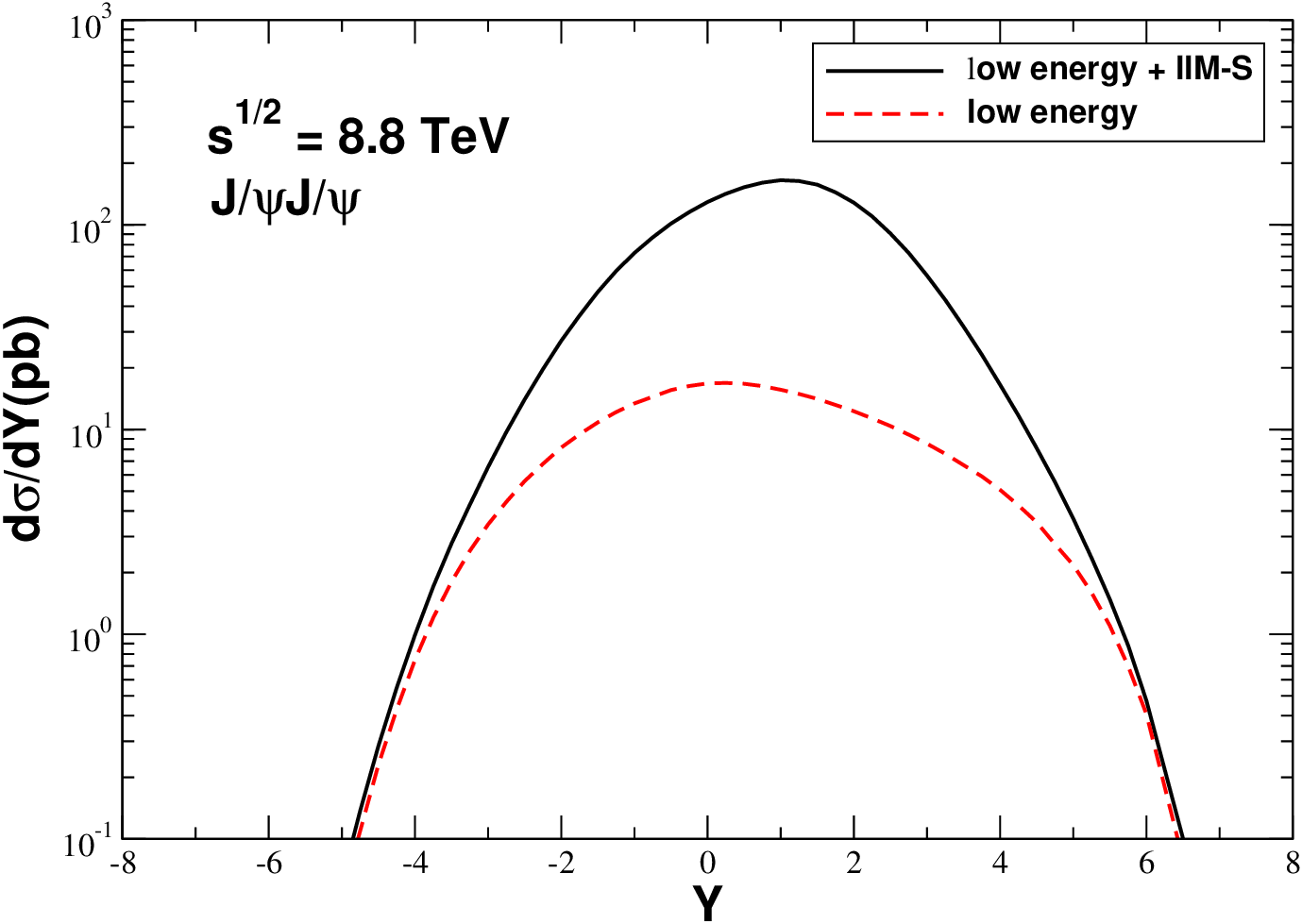,width=7cm}}
\end{tabular}                                                                                                                       
\caption{Rapidity distribution in double vector meson production in $\gamma \gamma$ interactions at $pPb$ collisions considering different 
values of $\sqrt{s}$.}
\label{fig3}
\end{figure}

\begin{figure}
\begin{tabular}{ccc}
{\psfig{figure=pp-rho-rho-500.eps,width=5cm}} & 
{\psfig{figure=pp-rho-rho-7000.eps,width=5cm}} & 
{\psfig{figure=pp-rho-rho-14000.eps,width=5cm}} \\
{\psfig{figure=pp-psi-psi-500.eps,width=5cm}} & 
{\psfig{figure=pp-psi-psi-7000.eps,width=5cm}} & 
{\psfig{figure=pp-psi-psi-14000.eps,width=5cm}}
\end{tabular}                                                                                                                       
\caption{Rapidity distribution for double vector meson production in $\gamma \gamma$ interactions at $pp$ collisions considering different 
values of $\sqrt{s}$.}
\label{fig4}
\end{figure}

\begin{figure}[t]
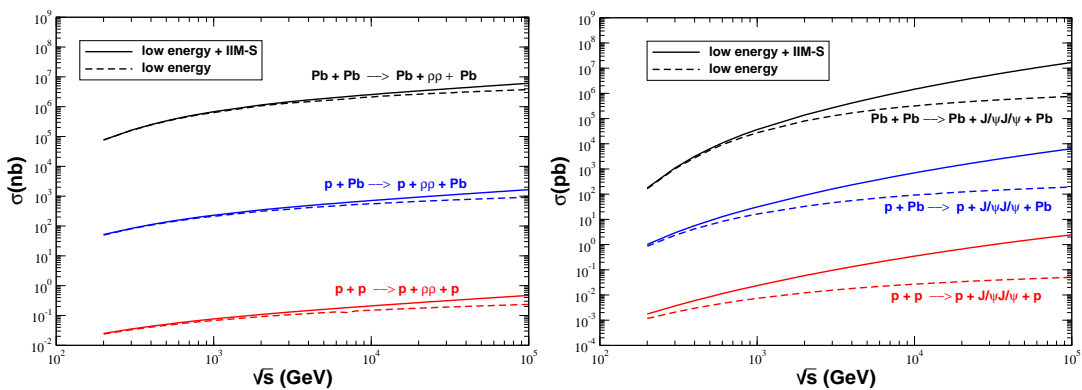

\begin{tabular}{cc}
{\psfig{figure=rho-rho-energia.eps,width=7cm}} & {\psfig{figure=psi-psi-energia.eps,width=7cm}} 
\end{tabular}                                                                                                                       
\caption{Energy dependence of the total cross section of double vector meson production in $\gamma \gamma$ interactions at $pp$, $pPb$ and $PbPb$ collisions.}
\label{fig5}
\end{figure}

In what follows we present our predictions for the rapidity distributions and total cross sections for  $\rho- \rho$ and $J/\Psi - J/\Psi$ production through  
$\gamma \gamma$ interactions in $pp$, $pPb$ and $PbPb$ collisions. In order to estimate the contribution of the gluonic part associated with the dipole - dipole 
interaction, 
we will compare the full predictions, obtained considering the low and high energy contributions (denoted Low energy + IIM-S hereafter) with those without the 
high energy contribution (Low energy hereafter). In order to estimate the equivalent photon spectra for $A = Pb$, we will consider the a monopole form factor 
$F(q^2) = \Lambda^2/(\Lambda^2 + q^2)$, with $\Lambda = 0.088$ GeV adjusted to reproduce the root - mean - square (rms) radius of the nucleus. Moreover, we will 
assume that $R_A = 1.2 \,  A^{\frac{1}{3}}$ fm. In the proton case, we will consider that   
$F(q^2) = 1/[1 + q^2/(0.71 \mbox{GeV}^2)]^2$ and $R_p = 0.7$ fm. Finally, as in \cite{vicmag07} we will assume $B_{\rho \rho} = 10$ GeV$^{-2}$ and 
$B_{\psi \psi} = 0.44$ GeV$^{-2}$.
Our results for the rapidity distributions are presented in Figs. \ref{fig2} -- \ref{fig4}. In the case of $Pb Pb$ collisions, presented  in Fig. \ref{fig2}, 
we obtain that the Low energy and Low Energy + IIM-S predictions are almost identical for  double $\rho$ production for all considered energies, which 
indicates that the gluonic contribution for this process is very small, in agreement with the conclusion obtained in Ref. \cite{antonirhorho}. This result 
is can be related  to the energy behavior of the $\gamma \gamma \rightarrow \rho  \rho$ cross section at high energies, which presents a mild growth  
with $W$, expected in a process dominated by large dipoles, and to the fact that in $AA$ collisions we are probing values of $W \le 160$ GeV for 
$\sqrt{s} = 5.5$ TeV. In contrast, for  double $J/\Psi$ production, we observe that the gluonic contribution increases with the energy, which is associated 
to the steep energy behavior of the $\gamma \gamma \rightarrow J/\Psi J/\Psi$ cross section. For $\sqrt{s} = 0.5$ TeV the analysis of this process  
can be useful to probe the box mechanism. On the other hand, for $\sqrt{s} = 5.5$ TeV the gluonic contribution implies an enhancement by a factor 2 of the 
rapidity distribution at $Y = 0$.  When $pPb$ collisions are considered, we obtain the asymmetric rapidity distributions presented in Fig. \ref{fig3}, which is 
expected since the nuclear equivalent photon spectra is enhanced by a factor $Z^2$. Moreover, in this case the energy range probed in the $\gamma \gamma$ interactions 
increases for $W \le 160$  GeV for $\sqrt{s} = 8.8$ TeV. As a consequence, we observe that the difference between the Low energy and Low Energy + IIM-S predictions 
starts to differ in the case of  double $\rho$ production and becomes appreciable for  double  $J/\Psi$ production. Finally, in Fig. \ref{fig4} we present our results 
for $pp$ collisions for different values of $\sqrt{s}$. In this case the  double vector meson production is induced by $\gamma \gamma$ interactions with 
$W \le 4500$ GeV for $\sqrt{s} = 14$ TeV. This large range of energies probed in the interaction implies that the gluonic contribution becomes very important for 
the description of  double vector production, increasing the rapidity distributions at $Y = 0$ by a factor 2 in the case of  $\rho \rho$ production and by a factor 40 
for  double $J/\Psi$ production in $pp$ collisions at 14 TeV. 
The importance of the gluonic contribution can also be estimated by the analysis of the energy dependence of the total cross section for  double vector meson
production in $\gamma \gamma$ interactions at $pp$, $pPb$ and $PbPb$ collisions. Our results are presented in Fig. \ref{fig5}. In agreement with our previous discussion, 
we can see that this contribution is small for  double $\rho$ production and appreciable for  double $J/\Psi$ production, mainly in $pp$ collisions. This result 
indicates that the analysis of  double $J/\Psi$ production in ultra peripheral hadronic collisions can be useful to study the QCD dynamics at high energies, as originally   
suggested in Ref.  \cite{vicmagvv1}.  In Tables \ref{tab:rho} and \ref{tab:jpsi} we present our predictions for total cross sections for  double vector meson production 
in $pp$, $pPb$ and $PbPb$ collisions for the energies of RHIC and LHC as well as for the  conceptual design energies  of the FCC \cite{fcc} and  CEPC - SPPC \cite{cepc}. 
It is important to emphasize that our Low energy + IIM-S predictions can be considered as a lower bound for the gluonic contribution, since other models for 
the dipole -  dipole cross section or for the description of the QCD dynamics imply larger values for the  
$\gamma \gamma \rightarrow J/\Psi J/\Psi$ cross section (For a detailed discussion see Ref. \cite{brunodouble}). Consequently, we believe that the analysis of 
this process is feasible in hadronic colliders. Additionally, considering the results from Ref. \cite{brunodouble} that  indicate that 
$\gamma \gamma \rightarrow V_1 V_2$ cross sections for the  
$\rho J/\Psi$, $\phi J/\Psi$, $\rho \Upsilon$, $J\Psi \Upsilon$ and $\Upsilon\Upsilon$ 
production  increase strongly with the energy, we  can also expect that these final states could be analysed in the future. As discussed in detail in 
Refs. \cite{vicmagvv1,vicmagvv2,vicmagvv3,brunodouble}, the study of these different final states is important to understand the transition between the soft 
and hard regimes of the QCD dynamics, since different dipole sizes are probed in each process.

\begin{table*}[t]
  \centering
  \begin{tabular}    {p{0.35\linewidth}p{0.25\linewidth}p{0.25\linewidth}}
    \hline
     \hline
        & Low energy & Low energy + IIM-S \\
    \hline
    \hline
    $PbPb \,\, (\sqrt{s}=500\,\mbox{GeV})$  & 0.33 $\times 10^{6}$  &  0.33 $\times 10^{6}$ \\ 
    $PbPb \,\, (\sqrt{s}=2.76\,\mbox{TeV})$ & 1.27 $\times 10^{6}$  & 1.39 $\times 10^{6}$ \\  
    $PbPb \,\, (\sqrt{s}=5.5\,\mbox{TeV})$ & 1.73 $\times 10^{6}$  & 1.97 $\times 10^{6}$ \\ 
    $PbPb \,\, (\sqrt{s}=39\,\mbox{TeV})$ &  3.11 $\times 10^{6}$  &  4.35 $\times 10^{6}$ \\ \hline
    $pPb \,\, (\sqrt{s}=5\,\mbox{TeV})$  & 449.45  & 536.43   \\ 
    $pPb \,\, (\sqrt{s}=8.8\,\mbox{TeV})$ & 535.32  & 678.46  \\  
    $pPb \,\, (\sqrt{s}=63\,\mbox{TeV})$  & 851.82  & 1408.95   \\ \hline
    $pp \,\, (\sqrt{s}=500\,\mbox{GeV})$  & 0.047  & 0.051  \\  
    $pp \,\, (\sqrt{s}=7\,\mbox{TeV})$  & 0.14  & 0.18  \\ 
    $pp \,\, (\sqrt{s}=13\,\mbox{TeV})$ & 0.16  & 0.23  \\ 
    $pp \,\, (\sqrt{s}=14\,\mbox{TeV})$ & 0.17  & 0.24  \\  
    $pp \,\, (\sqrt{s}=100\,\mbox{TeV})$ & 0.24  & 0.47  \\  
    \hline
       \hline
  \end{tabular}
  \caption{Total cross sections for  double $\rho$ production in $\gamma \gamma$ interactions at $pp$, $pPb$ and $PbPb$ collisions for  RHIC, LHC, FCC and 
CEPC - SPPC energies. Values em nb.}
  \label{tab:rho}
\end{table*}

\begin{table*}[t]
  \centering
  \begin{tabular}    {p{0.35\linewidth}p{0.25\linewidth}p{0.25\linewidth}}
    \hline
     \hline
        & Low energy & Low energy + IIM-S \\
    \hline
    \hline
    $PbPb \,\, (\sqrt{s}=500\,\mbox{GeV})$  & 5640  &  6423 \\ 
    $PbPb \,\, (\sqrt{s}=2.76\,\mbox{TeV})$ & 116550  & 235565 \\  
    $PbPb \,\, (\sqrt{s}=5.5\,\mbox{TeV})$ & 217019  & 658589 \\ 
    $PbPb \,\, (\sqrt{s}=39\,\mbox{TeV})$ & 578195  & 6861251 \\ \hline
    $pPb \,\, (\sqrt{s}=5\,\mbox{TeV})$  & 64  & 310  \\ 
    $pPb \,\, (\sqrt{s}=8.8\,\mbox{TeV})$ & 86  & 607  \\  
    $pPb \,\, (\sqrt{s}=63\,\mbox{TeV})$  & 172  & 4309   \\ \hline
    $pp \,\, (\sqrt{s}=500\,\mbox{GeV})$  & 0.0038  & 0.0085  \\  
    $pp \,\, (\sqrt{s}=7\,\mbox{TeV})$  & 0.023  & 0.24  \\ 
    $pp \,\, (\sqrt{s}=13\,\mbox{TeV})$ & 0.029  & 0.45  \\ 
    $pp \,\, (\sqrt{s}=14\,\mbox{TeV})$ & 0.030  & 0.48  \\  
    $pp \,\, (\sqrt{s}=100\,\mbox{TeV})$ & 0.050 & 2.42  \\
    \hline
     \hline 
  \end{tabular}
  \caption{Total cross sections for  double $J/\psi$ production in $\gamma \gamma$ interactions at at $pp$, $pPb$ and $PbPb$ collisions for RHIC, LHC, FCC 
and CEPC - SPPC energies. Values em pb.}
  \label{tab:jpsi}
\end{table*}

Finally, let us summarize our main conclusions. In recent years, a series of studies have discussed in detail the computation of the total cross section and 
the exclusive production of different final states in $\gamma \gamma$ interactions considering very distinct theoretical approaches. One of the basic motivations 
for these efforts is the possibility to study the behavior of  QCD dynamics at high energies. The ideal laboratory for these studies is the scattering of two 
off-shell photons at high energy in $e^+\,e^-$ colliders, which could be performed in the International Linear Collider (ILC). However, as the schedule for the 
construction and operation of this collider  is still an open question, the analysis of alternative ways to study the $\gamma \gamma$ interactions is an important 
theme. The  study of  double vector meson production in  $\gamma \gamma$ interactions in ultra peripheral heavy ion collisions as a probe of the QCD dynamics was 
proposed in Ref. \cite{vicmagvv1} and developed in Refs. \cite{vicmagvv2,vicmagvv3}. However, these studies  focused only on the high energy regime and disregarded 
the low energy mechanisms for  double vector production. As emphasized in Refs. \cite{antonirhorho,antonipsipsi}, the contribution of these mechanisms is important 
in $AA$ collisions, since the maximum center of mass energies probed in the $\gamma \gamma$ interactions is not large and the main contribution of  the equivalent 
photon spectrum comes from photons with low energy. However, these studies have disregarded the effects of the QCD dynamics discussed in    
Refs. \cite{vicmagvv2,vicmagvv3} and recently updated in Ref. \cite{brunodouble}. In this paper we have combined these two approaches and derived  predictions for 
the $\gamma \gamma \rightarrow VV$ cross section which are valid in the full kinematical range. We have obtained realistic predictions for the total cross sections 
in hadronic collisions and estimated the relative contribution of the low and high energy regimes. In particular, the results for $pp$ and $pPb$ have been derived 
by the first time. Our results demonstrated that   double $\rho$ production  is dominated by  low energy mechanisms. On the other hand, the gluonic contribution 
for double $J/\Psi$ production strongly increases with the energy,  the study of this process becomes feasible in hadronic collisions (mainly in $pp$ collisions)  
and it  may be useful to constrain the QCD dynamics at high energies, as proposed originally in Ref. \cite{vicmagvv1}.

\begin{acknowledgments}

This work was  partially financed by the Brazilian funding agencies CNPq, CAPES, FAPERGS and FAPESP.

\end{acknowledgments}

\hspace{1.0cm}

\end{document}